\documentclass[aps,prl,reprint,showpacs,groupedaddress]{revtex4-1}
\usepackage{amstext}
\usepackage{graphicx}
\usepackage{dcolumn}
\usepackage{bm}
\usepackage{subfigure}

\begin{document}


\title{Resolution of Single Spin-Flips of a Single Proton}

\author{A. Mooser,$^{1,2}$ H. Kracke,$^{1,2}$ K. Blaum,$^{3,4}$ S.$\,$A. Br\"auninger,$^{3,4}$ K. Franke,$^{3,5}$\\ C. Leiteritz,$^1$ W. Quint,$^{4,6}$ C.$\,$C. Rodegheri,$^{1,3}$ S. Ulmer,$^{5}$ J. Walz,$^{1,2}$}
\affiliation{$^1$ Institut f\"ur Physik, Johannes Gutenberg-Universit\"at D-55099 Mainz, Germany}
\affiliation{$^2$ Helmholtz-Institut Mainz, D-55099 Mainz, Germany}
\affiliation{$^3$ Max-Planck-Institut f\"ur Kernphysik, Saupfercheckweg 1, D-69117 Heidelberg, Germany}
\affiliation{$^4$ Ruprecht Karls-Universit\"at Heidelberg, D-69047 Heidelberg, Germany}
\affiliation{$^5$ RIKEN Advanced Science Institute, Hirosawa, Wako, Saitama 351-0198, Japan}
\affiliation{$^6$ GSI - Helmholtzzentrum f\"ur Schwerionenforschung, D-64291 Darmstadt, Germany}

\date{\today}

\begin{abstract}
The spin magnetic moment of a single proton in a cryogenic Penning trap was coupled to the particle's axial motion with a superimposed magnetic bottle. Jumps in the oscillation frequency indicate spin-flips and were identified using a Bayesian analysis.
\end{abstract}

\pacs{14.20.Dh 21.10.Ky 37.10.Ty}

\maketitle
Recent dramatic advances in quantum control of a single isolated nucleus have opened the way for direct precision measurements of the proton and antiproton magnetic moments. The present most precise value for the proton magnetic moment comes from measurements of the hyperfine splitting in atomic hydrogen \cite{winkler1972magnetic}. Bound-state corrections have to be included to extract the magnetic moment of the free proton with a precision of $8.2$ parts in $10^{9}$ \cite{Mohr}. The antiproton magnetic moment has been determined with a precision of $2$ parts in $10^{3}$ \cite{PartData} from super-hyperfine spectroscopy of antiprotonic helium \cite{Pask} and from fine-structure spectroscopy of antiprotonic lead \cite{kreissl}.\\
A direct precision measurement with just one isolated particle in a Penning trap has the potential to improve the precision of the value of the proton magnetic moment $\mu_p$ by more than one order of magnitude. In addition, there would be no need for theoretical corrections. Thus a direct measurement could be used for consistency tests with previous measurements based on hyperfine splittings. In the case of the antiproton the potential for improvement is more than six orders of magnitude. This would enable another stringent test of the symmetry between matter and antimatter in the baryon sector \cite{Bluhm}.\\
The principle of a direct measurement of $\mu_p$ is to determine the Larmor (spin-precession) frequency $\nu_L$ and the cyclotron frequency $\nu_c$ of a proton in a magnetic field. The frequency ratio $\nu_L/\nu_c=\mu_p/\mu_N$ gives the proton magnetic moment $\mu_p$ in terms of the nuclear magneton $\mu_N$. The cyclotron frequency $\nu_c$ of a proton in a Penning trap can readily be measured by preparing a double-dressed state \cite{Ulmer2} and applying the Brown-Gabrielse invariance theorem \cite{Brown22}. The Larmor frequency $\nu_L$ can be determined by driving spin-flips and measuring the transition probability as a function of the drive frequency. To this end, an inhomogeneous magnetic field, a ``magnetic bottle,'' is used which couples the spin magnetic moment to the axial motion of the proton. Using this ``continuous Stern-Gerlach effect'' \cite{Dehmelt,Quint}, jumps in the axial oscillation frequency indicate spin-flips. The challenge is to detect these spin-flips on a background of axial frequency changes which result from tiny changes of the motional angular momentum of the proton in the trap.\\
Recently we reported on the statistical detection of spin-flips of a single proton in an inhomogeneous magnetic field \cite{Ulmer}. We have used this method to obtain $8.9\cdot10^{-6}$ \cite{CCR2}, while a similar experiment by another group achieved $2.5\cdot10^{-6}$ \cite{DiSciacca}. Both measurements are limited by the inhomogeneity of the magnetic field. The precision can be boosted by several orders of magnitude using the double-Penning trap technique \cite{Haeffner}. In this elegant method the measurement of $\nu_c$ and the excitation of spin-flips at $\nu_L$ happens in a first Penning trap with a homogeneous magnetic field. Spin-state detection is carried out in a second Penning trap with a magnetic bottle. This method has been used to measure the magnetic moment of the electron bound in $^{28}$Si$^{13+}$ with a relative precision of 5$\cdot$10$^{-10}$ \cite{Sven}. The double-Penning trap technique requires that single spin-flips can be resolved, which so far was not possible with nuclear spins.\\ 
In this Letter, we present the first detection of \textit{single} spin-flips of a single proton. Noise-driven random transitions between the cyclotron quantum states cause a background of frequency fluctuations. The characterization of these frequency fluctuations enables a novel spin state analysis method for Penning trap experiments based on a Bayesian formalism.\\
\begin{figure}[t]
        \centerline{\includegraphics[width=8cm,keepaspectratio]{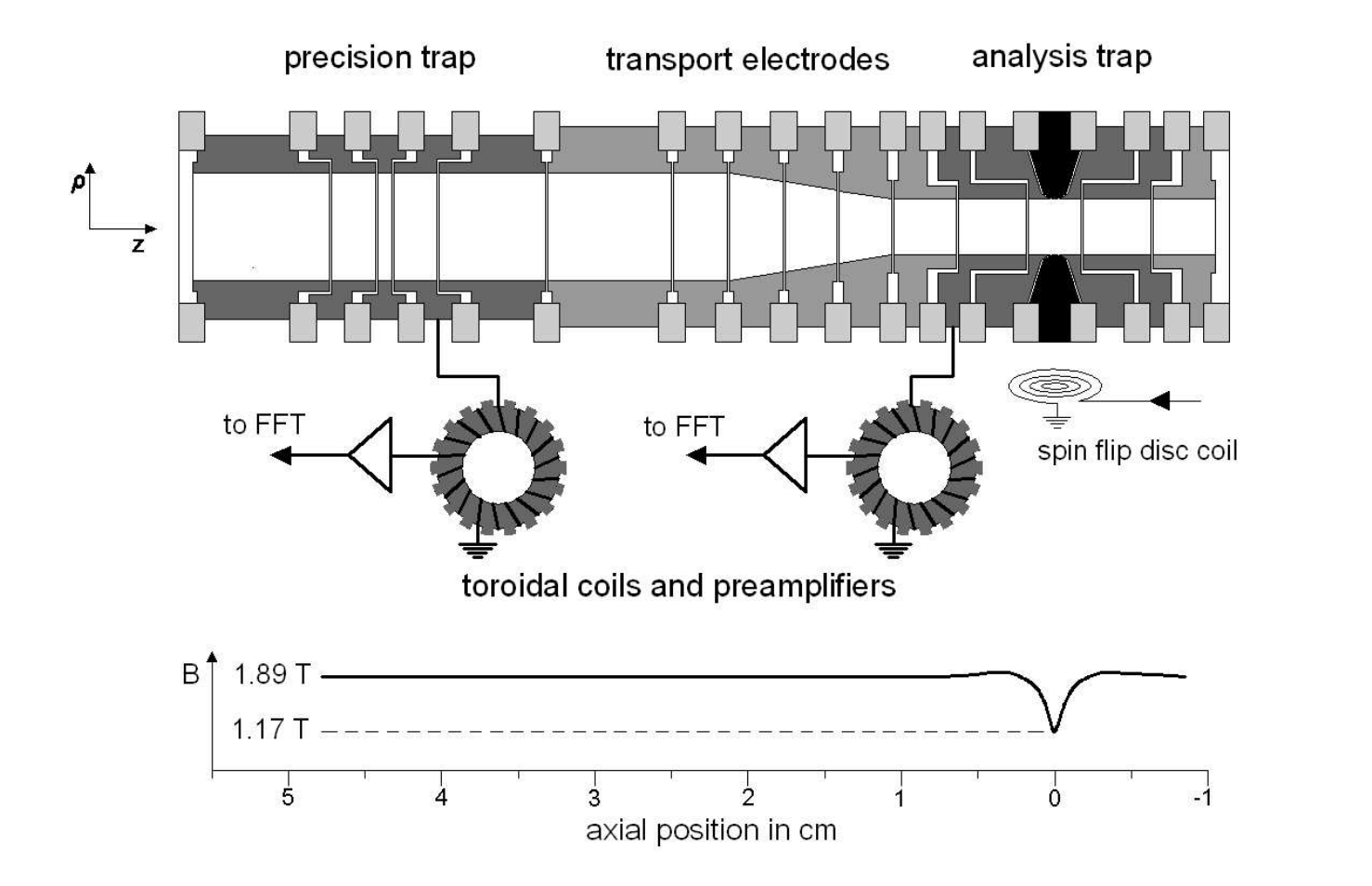}}
            \caption[FEP]{Schematic of the experiment. Two cylindrical Penning traps are connected by transport electrodes. Superconducting toroidal coils and cryogenic low-noise preamplifiers are used for the detection of the axial motion. The signals are analyzed by a fast Fourier transform, FFT. The central ring electrode of the analysis trap (black) is made of a ferromagnetic Co/Fe alloy. The magnetic field along the \textit{z} axis is indicated in the lower graph. Voltages are applied to the electrodes using low-pass filters (not shown). For further details see text.}
\label{fig:TrapDetect}
\end{figure}
Our apparatus consists of a cryogenic double-Penning trap, mounted in a superconducting magnet and cooled by a liquid helium cryostat. Both Penning traps are shown in Fig.\ \ref{fig:TrapDetect} and have five electrodes in compensated and orthogonal design \cite{CCR2,gabrielse1989oep}. The \emph{precision trap} is placed in the central homogeneous volume of the $1{.}89\,$T magnetic field. The \emph{analysis trap} has a ring electrode made of ferromagnetic Co/Fe material, which shapes the magnetic field to a so-called magnetic bottle
\begin{equation}
\vec{B}\left(z,\rho\right)=B_0\hat{e}_z+B_2\left[\left(z^2-\frac{\rho^2}{2}\right)\hat{e}_z-z\rho\hat{e}_{\rho}\right],
\end{equation}
with $B_2=2.97(10)\cdot10^5\,$T$/$m$^{2}$. The proton can be moved between both traps using transport electrodes. The whole electrode stack is placed in a sealed vacuum chamber and cryopumping is utilized. Collisions with residual gas are negligible and single particles can be trapped for months.\\
In the Penning trap the proton has three eigenmotions, the axial motion with frequency $\nu_z$, and two radial motions: the magnetron and the modified cyclotron motion, at frequency $\nu_-$ and $\nu_+$, respectively. The eigenfrequencies are $\nu_z=623\,$kHz, $\nu_-=8\,$kHz, $\nu_+=28.9\,$MHz in the precision trap, and $\nu_z=742\,$kHz, $\nu_-=15\,$kHz, $\nu_+=17.9\,$MHz in the analysis trap. The motion of the proton induces image currents in the trap electrodes. These currents in the \textit{fA} range are measured by connecting a superconducting inductance $L$ to the trap, which forms a resonant circuit together with the parasitic trap capacitance. The resonant circuit has a quality factor $Q$ and acts as an effective parallel resistance $R_p=2\pi\nu_{res}QL$ on its resonance frequency $\nu_{res}$. The signals are amplified and a fast Fourier transform (FFT) is performed to access the particle's eigenfrequency. Both traps are connected to detection circuits for the axial motion. The detection system for the precision trap has a quality factor of 12500, resulting in $R_p=130\,$M$\Omega$ and a signal-to-noise ratio of $S/N=20\,$dB. Here the noise is given by $N=u_{n}$, with $u_n$ the voltage noise density of the preamplifier, and the signal by $S^2=4k_BTR_p\Delta\nu\alpha^2+ i^2_n R^2_p\alpha^4 + u^2_{n}$, with $i_n$ the current noise density of the preamplifier, $\Delta\nu$ the bandwidth and $\alpha$ the decoupling of the inductor and preamplifier \cite{jeff}. The resonator for the analysis trap has a quality factor of 9500 with $R_p=85\,$M$\Omega$ at $S/N=25\,$dB. At $\nu_{res}=\nu_{z}$ the trapped proton shorts the thermal noise of the detector. This causes a dip in the noise spectrum of the detector from which $\nu_z$ can be determined \cite{Feng}. Additionally, one cyclotron damping coil is connected to a split electrode of the precision trap \cite{Ulmer3}.\\
The proton spin magnetic moment $\mu_p$ is coupled to the axial motion by the magnetic bottle in the analysis trap. This causes an additional potential term $\Phi_{mag}=-\vec{\mu}\cdot\vec{B}$, which adds to the electric quadrupolar potential $\Phi_E$ of the Penning trap. $\vec{\mu}$ is the sum of the magnetic moments $\vec{\mu}_+$ and $\vec{\mu}_-$ due to the modified cyclotron- and magnetron motion, respectively, and the spin magnetic moment $\vec{\mu}_p$. The total axial potential energy of the single particle is $E_{pot}=q_p\Phi_E-\vec{\mu}\vec{B}$ and the axial frequency is
\begin{eqnarray}
\nu_z &=& \sqrt{\frac{1}{m_p}\left.\frac{\partial^2E_{pot}}{\partial z^2}\right|_{z=0}}\nonumber \\
      &=& \sqrt{\frac{2e C_2 V_0}{m_p}+\frac{2}{m_p}\left(\mu_p+\mu_++\mu_-\right)B_2}\,,
\end{eqnarray}
where $C_2$ is a geometry parameter of the trap and $V_0$ the trapping potential. Changes in $\mu_p B_0$, $E_+$ and $\left|E_-\right|$ shift the axial frequency by
\begin{equation}
\Delta\nu_z\approx\frac{1}{4\pi^2m_p\nu_z}\frac{B_2}{B_0}\left( E_++\left| E_-\right|\pm \mu_p B_0\right).
\label{Equ:dnuz}
\end{equation}
In our analysis trap a proton spin-flip causes an axial frequency jump of $\Delta\nu_{z,sf}=171\,$mHz at 742$\,$kHz. The strong magnetic bottle, however, also makes the axial frequency extremely sensitive to the energy in the radial modes. A change of only 4 parts in $10^{4}$ of the thermal cyclotron energy of $360\,\mu$eV ($4.2\,$K) causes the axial frequency to change by the same amount as a proton spin-flip.\\
\begin{figure}[h]
 \centering
  \includegraphics[width=0.48\textwidth]{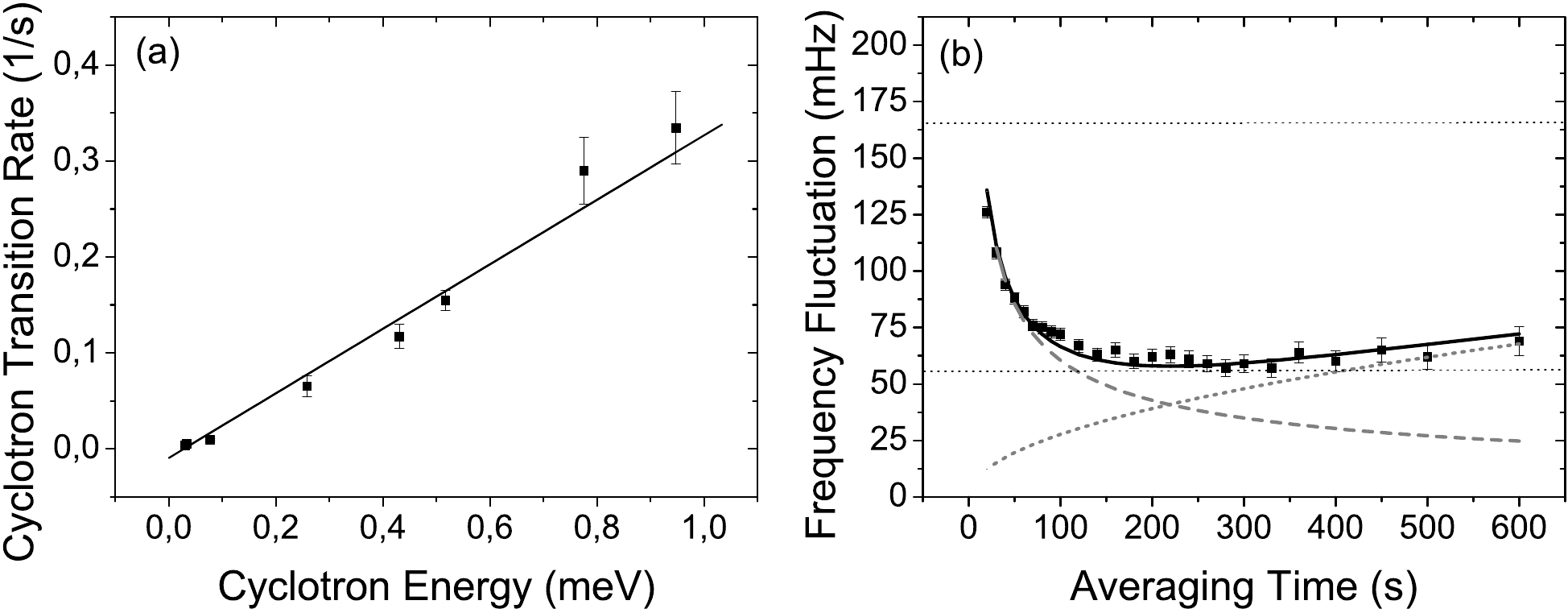}
  \caption{(a) Cyclotron transition rate as function of the cyclotron energy. (b) Measurement of the axial frequency stability as a function of averaging time. The dashed line which decreases is due to detector white noise averaging. The dotted line comes from a calculation with white noise driving the cyclotron mode (see text). The lower horizontal line indicates the best stability of $\Xi_{opt}=55\,$mHz achieved. The upper horizontal line indicates the size of the frequency jump due to a spin-flip of $171\,$mHz, which is about $3\Xi_{opt}$.}
 \label{fig:StabilityAT}
\end{figure}
Therefore, for the detection of individual spin-flips the stability of the radial energy is crucial. Transitions between the cyclotron quantum states $n_+$ cause axial frequency jumps of $\Delta\nu_{z,+}=\pm63\,$mHz. As a result the axial frequency fluctuates with $\Xi^2_{cyc}=(\delta n_+/\delta t) T \Delta\nu^2_{z,+}$, where $T$ is the FFT averaging time. The cyclotron transition rate can be calculated using Fermi's golden rule
    \begin{equation}
    \frac{\delta n_+}{\delta t} =\frac{2\pi}{\hbar}\Delta_+\rho(E_+)\Gamma^2_{i\rightarrow f}.
    \label{Equ:Fermi}
    \end{equation}
Here $\rho(E_+)$ is the density of states of the 1-dimensional harmonic oscillator and $2\pi\Delta_+=e B_2\left\langle z^2\right\rangle /m_\text{p}$ is the line-width of the cyclotron resonance \cite{Brown33} due to the coupling of the axial mode to the thermal bath of the detection system. $\left\langle z^2\right\rangle$ is the expectation value of the square of the axial amplitude.
    \begin{equation}
  \Gamma_{i\rightarrow f}=qE_0\sqrt{\frac{\hbar}{m_\text{p} \omega_+}\frac{n_+}{2}}
    \end{equation}
is the electric dipole cyclotron transition matrix element with $E_0$ the spectral noise density of a spurious electric field. Note that Eq.\ (\ref{Equ:Fermi}) gives a cyclotron transition rate which is proportional to the cyclotron energy, $\delta n_+/\delta t\propto n_+$.\\ 
In the experiment $E_+$ was calibrated by determining $\nu_z+\Delta\nu_z(E_+)$ with the method described in \cite{Djekic}. The proton's cyclotron mode was thermalized with the damping coil in the precision trap several times, which results in Boltzmann distributed frequency shifts $\Delta\nu_z(E_+)$ in the analysis trap. $E_+=0$ is then identified by $\Delta\nu_z=0$. After the calibration $\Xi_{cyc}$ was measured for different cyclotron energies and the transition rate $\delta n_+/\delta t$ was determined as a function of $E_+$. The measurement data are shown in Fig.\ \ref{fig:StabilityAT}(a) and are consistent with a linear relation as predicted above. $(\delta n_+/\delta t) /E_+=0.35\,$meV$^{-1}$s$^{-1}$ was obtained, which corresponds to a noise drive of $E_0=$7$\,$nV$\cdot$m$^{-1}\cdot$Hz$^{-1/2}$. To obtain $\Xi_{cyc}<\Delta\nu_{z,sf}/3$ at a typical FFT averaging time of $200\,$s a proton with a cyclotron energy of less than $10\,\mu$eV has to be selected.\\
\begin{figure}[ht]
 \centering
  \includegraphics[width=0.48\textwidth]{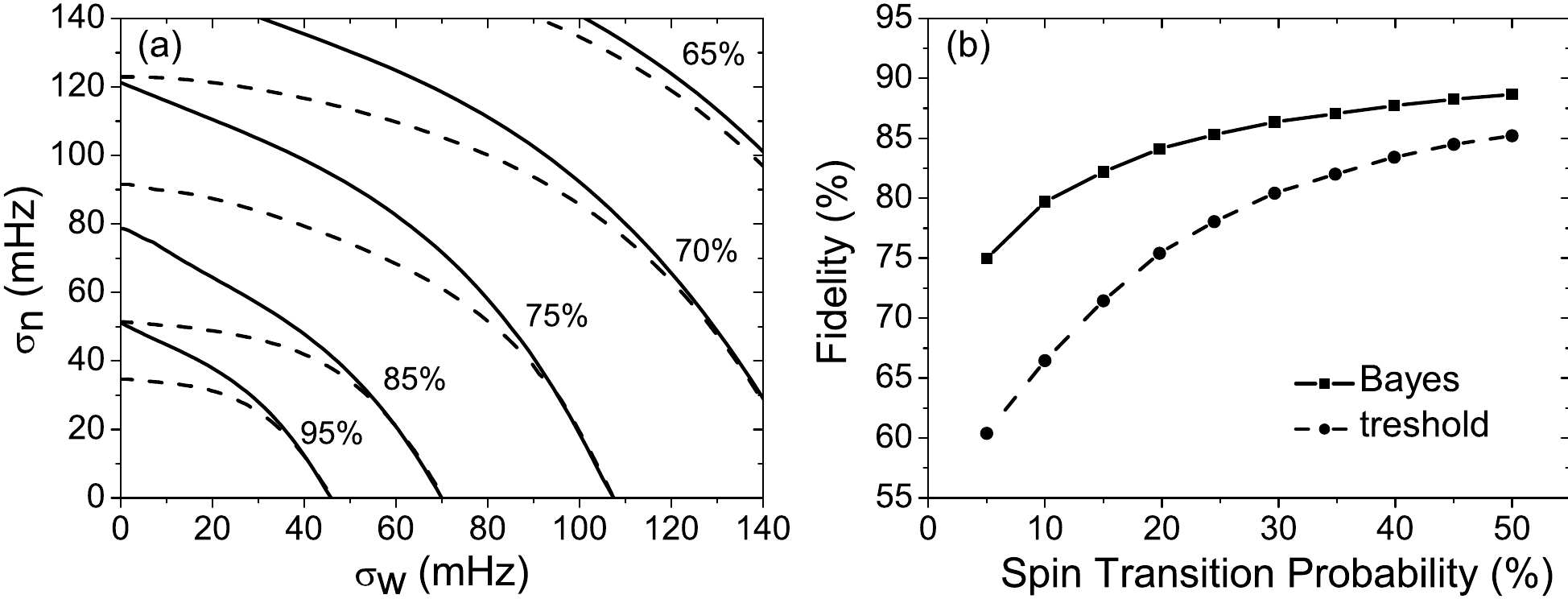}
\caption{Comparison of the threshold method and Bayes analysis. (a) shows contour lines of the fidelity for a spin-flip probability of 50\% as a function of the white noise $\sigma_n$ and random walk $\sigma_w$ contribution; solid lines: Bayes analysis, dashed lines: threshold method. (b) shows the fidelity as a function of the spin-flip probability at $\sigma_w=\sigma_n=39\,$mHz.}
\label{fig:bayes}
\end{figure}
As a measure for the axial frequency fluctuation we define $\Xi$ as the standard deviation of the difference between two subsequent axial frequency measurements $\alpha(T)=\nu_z(t)-\nu_z(t+T)$, $\Xi(T)=\sqrt{(N-1)^{-1}\sum_N(\alpha(T)-\bar{\alpha}(T))^2}$. $\Xi$ is invariant under drifts and, in the case of $\left\langle \alpha(T)\right\rangle=0$, has the same characteristics as the common Allan deviation \cite{nist}. In Fig.\ \ref{fig:StabilityAT}(b) the variation of $\Xi$ with the measuring time $T$ is shown. For short measuring times $\Xi$ decreases with $1/\sqrt{T}$ due to averaging of the detector white noise as indicated by the dashed line. At measuring times above 300$\,$s $\Xi$ increases with $\sqrt{T}$, which is due to a random walk caused by fluctuations of the cyclotron energy. The dotted line is the result of a calculation with a transition rate for cyclotron quantum jumps of $\delta n_+/\delta t=0.002\,$s$^{-1}$.\\
The lowest axial frequency fluctuation achieved is $\Xi_{opt}=55\,$mHz, which corresponds to a reduction of the frequency fluctuations due to the random walk and white noise by about 90$\,\%$ compared to our previous work \cite{CCR2}. This improvement is due to the detection system with a higher quality factor reducing the white noise contribution. By an increase of the effective electrode distance a stronger decoupling \cite{Feng} is possible. Because of the smaller line-width of the noise dip and the higher signal-to-noise ratio of the detection system the axial frequency is measured faster and more precisely. For a reduction of a cyclotron noise drive a split electrode of the analysis trap, initially used for coupling of the axial mode to radial modes, was replaced.\\
\begin{figure*}[t!]
  \centering
  \includegraphics[width=0.8\textwidth]{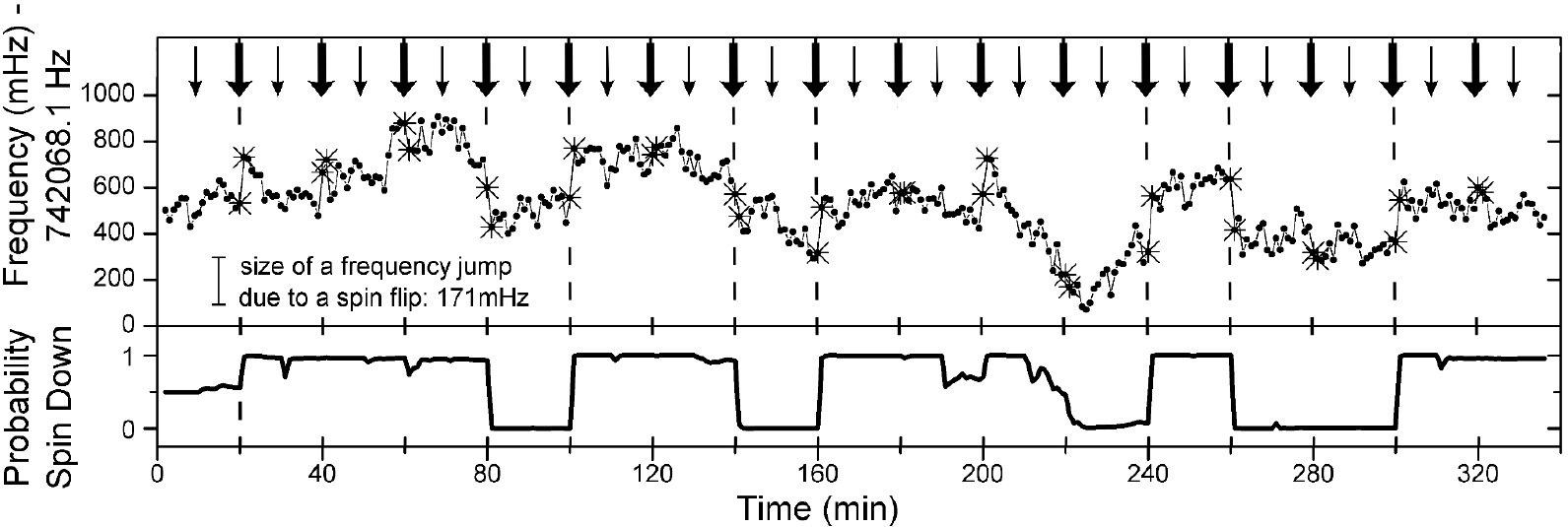}
   \caption{\label{fig:single} Observation of single spin-flips with a single proton. At the top a series of axial frequency measurements is shown. A resonant spin-flip drive was applied between the crossed data points, indicated by thick arrows.  In between an off-resonant spin-flip drive was turned on indicated by the thin arrows. Otherwise the drive was turned off. Large axial frequency jumps after a resonant drive are due to single proton spin-flips. At the bottom the result of the Bayesian analysis is shown.}
\end{figure*}
Spin state analysis can be done with a simple method assigning each axial frequency jump above a given threshold to a spin-flip. We developed an alternate and more advanced method, which is based on probability theory and uses an update probability theorem, Bayes rule \cite{Sivia}. In a series of measurements, each axial frequency $f_i$ is modeled as the sum of an accumulated random walk $W_i=\sum w_i$, white noise $n_i$ and a frequency jump due to a spin-flip with $f_i=W_i+n_i\pm1/2\Delta\nu_{z,sf}$. The distributions of $w_i$ and $n_i$ can be described by normal distributions $N(\mu;\sigma)$ with mean $\mu$ and standard deviations $\sigma_w$ and $\sigma_n$, respectively. The conditional probability $P^i_{\alpha,W}=P\left(\alpha_i,W_i\right|f_i,f_{i-1},...)$ is defined, which is the probability of being in a spin state $\alpha_i=\left\{\uparrow_i,\downarrow_i\right\}$ and having an accumulated random walk $W_i$ given all frequency information $f_i,f_{i-1},...$ . This definition allows the application of Bayes's rule relating the \textit{a-posteriori} probability $P^i_{\alpha,W}$ to the \textit{a-priori} probability $P\left(\alpha_i,W_i\right|f_{i-1},...)$,
\begin{equation}
   \label{Equ:bayes1}
    P^i_{\alpha,W} 
    =\frac{P\left(f_i|\alpha_i,W_i,f_{i-1},...\right) P\left(\alpha_i,W_i|f_{i-1},...\right)}{P\left(f_i|f_{i-1},...\right)},
\end{equation}
with $P\left(f_i|f_{i-1},...\right)$ the normalization found by $\int_{W}\sum_{\alpha} P^i_{\alpha,W}=1$. Due to the frequency correlation introduced by the random walk the conditional probability of measuring frequency $f_i$ after a measurement of $f_{i-1}$ cannot be regarded independently, $P(f_{i}|f_{i-1})\neq P(f_{i})$. In our definition $\alpha_i$ and $W_i$ completely specify the distribution of $f_i$ and the first factor in the numerator can be simplified, $P\left(f_i|\alpha_i,W_i,f_{i-1},...\right)=P\left(f_i|\alpha_i,W_i\right)$. Hence the first factor is the probability density for $f_i$, given the state as well as random walk information at time $i$. It depends on the white noise contribution,
\begin{eqnarray} 
  P(f_i|\alpha_i,W_i)&=& N(f_i-W_i\pm1/2\Delta\nu_{z,sf};\sigma_n),
\end{eqnarray} 
with $+$ for $\alpha_i=\uparrow_i$ and $-$ for $\alpha_i=\downarrow_i$, respectively. The second factor in the numerator of Eq.\ \ref{Equ:bayes1} depends on the previous state through the previous frequency measurements $f_{i-1},f_{i-2},...$ . Integrating over all possible states and random walks gives
\begin{eqnarray} 
	  P &(&\alpha_i, W_i |f_{i-1},...) \label{Equ:bayes2} \\
	  &=&\sum_{\alpha_{i-1}}\int_{W'}P\left(\alpha_i,W_i|\alpha_{i-1},W'_{i-1}f_{i-1},...\right)\times P^{i-1}_{\alpha,W}.\nonumber 
\end{eqnarray} 
The first factor describes the probability density without the knowledge of the latest frequency measurement. It introduces an update of the state probabilities describing the evolution of the state probability from measurement $i-1$ to $i$, 
\begin{eqnarray} 
  P&(&\alpha_i,W_i|\alpha_{i-1},W'_{i-1},f_{i-1},...)\nonumber \\
  &=& P(\alpha_i,W_i|\alpha_{i-1},W'_{i-1})\nonumber \\
  &=& (1-p_{sf})N(W_i-W'_{i-1};\sigma_w) \quad \mbox{for } \alpha_i=\alpha_{i-1}\\
  &=& p_{sf}N(W_i-W'_{i-1};\sigma_w) \quad \qquad \mbox{for } \alpha_i\neq\alpha_{i-1},
\end{eqnarray} 
with the spin transition probability $p_{sf}$. The second factor in Eq.\ (\ref{Equ:bayes2}) $P^{i-1}_{\alpha,W}$ simply is the state and random walk probability at time $i-1$. The final state probability at time i, $P^i_{\alpha}$, is obtained by usage of the marginalization rule $P^i_{\alpha}=\int_{W}P^i_{\alpha,W}$. Input parameters are the spin-flip probability $p_{sf}$, the random walk contribution $\sigma_w$ and the white noise contribution $\sigma_n$.\\
To investigate the quality of the spin state analysis we define the \textit{fidelity} as the fraction of correctly identified spin states in a series of frequency measurements. It was found numerically that the best fidelity in the threshold method is obtained for a threshold of $\Delta\nu_{z,sf}/2$.\\
Fig.\ \ref{fig:bayes}(a) shows the fidelity achieved with both, the threshold method, and the Bayes algorithm using simulated data. In the case of a pure random walk both methods give the same fidelity. If white noise contributes significantly the Bayes algorithm has better fidelity than the threshold method. In Fig.\ \ref{fig:bayes}(b) the fidelity is shown as a function of the spin-flip probability using again simulated data. The random walk and white noise contribution were set to values which correspond to the optimal frequency stability ($\sigma_w=\sigma_n$) in the experiment, $\Xi_{opt}=\sqrt{\sigma^2_w+\sigma^2_n}=55\,$mHz. The Bayes method clearly is superior to the threshold method and has a fidelity of 88$\,\%$ at 50$\,\%$ spin-flip probability.\\
The experimental observation of single spin-flips is shown in Fig.\ \ref{fig:single}. The upper trace shows a time-series of axial frequency measurements. Thin vertical arrows indicate times when an off-resonant radio-frequency was applied to the spin-flip coil (see Fig.\ \ref{fig:TrapDetect}). No extraordinary frequency jumps  are seen in the data at these times. This is an important background test which shows, that the spin-flip drive does not affect the cyclotron motion. Thick vertical arrows indicate times when a resonant drive was applied. Several axial frequency jumps whose size corresponds to spin-flips can clearly be seen in the raw data exactly at these times. One example of a spin-flip down and later up is at $240\,$min and $260\,$min. The lower trace in Fig.\ \ref{fig:single} is the result of the Bayesian spin-state analysis which is initialized with uncertainty (50$\,\%$ spin down probability) at $t=0$. Note, that the Bayesian analysis is causal \cite{Meschede}, i.e. at each point in time it uses only present and prior axial frequency data. The Bayesian analysis nicely confirms the spin-flips visible in the raw data and provides a consistent picture of the time-evolution of the proton spin-state projection.\\
In conclusion we observed single spin-flips of a single proton for the first time. This enables the application of the double Penning-trap method to measure magnetic moments of both the proton and the antiproton with $10^{-9}$ precision, or better.\\
We acknowledge fruitful discussions with Sven Sturm. J.$\,$W. acknowledges a helpful discussion with D. Meschede. This work was supported by the BMBF, the EU (ERC Grant No. 290870-MEFUCO), the Helmholtz-Gemeinschaft, the Max-Planck Society, the IMPRS-PTFS and the RIKEN Initiative Research Program.\\


\begin{thebibliography}{99}
\bibitem{winkler1972magnetic} P.$\,$F. Winkler, D. Kleppner, T. Myint, and F.$\,$G. Walther, Phys. Rev. A \textbf{5}, 83 (1972).
\bibitem{Mohr} P.$\,$J. Mohr, B.$\,$N. Taylor, and D.$\,$B. Newell, Rev. Mod. Phys. \textbf{80}, 633 (2008).
\bibitem{PartData} J. Beringer \textit{et al.} (Particle Data Group), Phys. Rev. D \textbf{86}, 010001 (2012).
\bibitem{Pask} T. Pask, D. Barna, A. Dax, R.$\,$S. Hayano, M. Hori, D. Horv\'ath, S. Friedreich, B. Juh\'asz, O. Massiczek, N. Ono, A. S\'ot\'er, and E. Widmann, Phys. Lett. B \textbf{678}, 55 (2009).
\bibitem{kreissl} A. Kreissl, A.$\,$D. Hancock, H. Koch, Th. K\"ohler, H. Poth, U. Raich, D. Rohmann, A. Wolf, L. Tauscher, A. Nilsson, M. Suffert, M. Chardalas, S. Dedoussis, H. Daniel, T. von Egidy, F.$\,$J. Hartmann, W. Kanert, H. Plendl, G. Schmidt, J.$\,$J. Reidy, Z. Phys. C Part. Fields \textbf{37}, 557 (1988).
\bibitem{Bluhm} R. Bluhm, V.$\,$A. Kostelecky, and N. Russell, Phys. Rev. D \textbf{57}, 3932 (1998).
\bibitem{Ulmer2} S. Ulmer, K. Blaum, H. Kracke, A. Mooser, W. Quint, C.$\,$C. Rodegheri, and J. Walz, Phys. Rev. Lett. \textbf{107}, 103002 (2011).
\bibitem{Brown22} L.$\,$S. Brown and G. Gabrielse, Phys. Rev. A \textbf{25}, 2423 (1982).
\bibitem{Dehmelt} H. Dehmelt, Proc. Natl. Acad. Sci. \textbf{83}, 2291 (1986).
\bibitem{Quint} W. Quint, J. Alonso, S. Djeki\'c, H.$\,$J. Kluge, S. Stahl, T. Valenzuela, J. Verd\'u, M. Vogel, and G. Werth, Nucl.
Instr. Meth. B \textbf{214}, 207 (2004).
\bibitem{Ulmer} S. Ulmer, C.$\,$C. Rodegheri, K. Blaum, H. Kracke, A. Mooser, W. Quint, and J. Walz,  Phys. Rev. Lett. \textbf{106}, 253001 (2011).
\bibitem{CCR2} C.$\,$C. Rodegheri, H. Kracke, K. Blaum, S. Kreim, A. Mooser, W. Quint, S. Ulmer, and J. Walz, New J. Phys. \textbf{14}, 063011 (2012).
\bibitem{DiSciacca} J. DiSciacca and G. Gabrielse, Phys. Rev. Lett. \textbf{108}, 153001 (2012).
\bibitem{Haeffner} H. H\"affner, T. Beier, N. Hermanspahn, H.-J. Kluge, W. Quint, S. Stahl, J. Verd\'u, and G. Werth, Phys. Rev. Lett. \textbf{85}, 5308 (2000).
\bibitem{Sven} S. Sturm, A. Wagner, B. Schabinger, J. Zatorski, Z. Harman, W. Quint, G. Werth, C.$\,$H. Keitel, and K. Blaum, Phys. Rev. Lett. \textbf{107}, 023002 (2011).
\bibitem{gabrielse1989oep} G. Gabrielse, L. Haarsma, and S.$\,$L. Rolston,  Int. J. Mass Spec. \textbf{88}, 319 (1989); \textit{ibid.} \textbf{93}, 121 (E) (1989).
\bibitem{jeff} S. Jefferts, T. Heavner, P. Hayes, and G.$\,$H. Dunn, Rev. Sci. Instrum. \textbf{64}, 737 (1993).
\bibitem{Feng} X. Feng, M. Charlton, M. Holzscheiter, R.$\,$A. Lewis, and Y. Yamazaki, J. Appl. Phys. \textbf{79}, 8 (1996).
\bibitem{Ulmer3} S. Ulmer, K. Blaum, H. Kracke, A. Mooser, W. Quint, C.$\,$C. Rodegheri, and J. Walz, accepted by Nucl. Instr. Meth. Phys. Res. A (2012).
\bibitem{Brown33} L.$\,$S. Brown, Ann. Phys. \textbf{159}, 62 (1985).
\bibitem{Djekic} S. Djekic, J. Alonso, H.-J. Kluge, W. Quint, S. Stahl, T. Valenzuela, J. Verd\'u, M. Vogel, and G. Werth, Eur. J. Phys. D \textbf{31}, 451 (2004).
\bibitem{nist} W.$\,$J. Riley, Natl. Inst. Stand. Technol. Spec. Publ. 1065 (2008).
\bibitem{Sivia} D.$\,$S. Sivia, \textit{Data Analysis: A Bayesian Tutorial} (Oxford University Press, Oxford, 1996).
\bibitem{Meschede} S. Brakhane, W. Alt, T. Kampschulte, M. Martinez-Dorantes, R. Reimann, S. Yoon, A. Widera, and D. Meschede, Phys. Rev. Lett. \textbf{109}, 173601 (2012).
\end{thebibliography}
\end{document}